\documentclass[aps,prl,preprint,groupedaddress]{revtex4-1}

\usepackage{graphicx}
\usepackage{color}
\usepackage{natbib}
\usepackage{amssymb}
\usepackage{amsmath}
\usepackage{amsfonts}
\begin{document}


\title{New insights into noninvasive imaging \\through strong scattering media}


\author{Yuan Yuan}
\thanks{These two authors contributed equally.}
\affiliation{Shaanxi Key Laboratory of Environment and Control for Flight Vehicle, Xi'an Jiaotong University, China,710049}

\author{Hui Chen}
\email[]{Corresponding author. chenhui@xjtu.edu.cn}
\affiliation{Electronic Material Research Laboratory, Key Laboratory of the Ministry of Education and International Centre for Dielectric Research, Xi'an Jiaotong University, China,710049}
\affiliation{These two authors contributed equally.}


\date{\today}

\begin{abstract}
    In turbid media, scattering of light scrambles information of the incident beam and represents an obstacle to optical imaging. Noninvasive imaging through opaque layers is challenging for dynamic and wide-field objects due to unreliable image reconstruction processes. We here propose a new perspective to solve these problems: rather than using the full point-spread-function (PSF), the wave distortions in scattering layers can be characterized with only the phase of the optical-transfer-function (OTF, the Fourier transform of PSF), with which diffraction-limit images can be analytically solved. We then develop a method that exploits the redundant information dynamic objects, and can reliably and rapidly recover OTFs’ phases within several iterations. It enables not only noninvasive video imaging at $25\sim 200\,Hz$ of a moving object hidden inside turbid media, but also imaging under weak illumination that is inaccessible with previous methods. Furthermore, by scanning a localized illumination on the object plane, we propose a wide-field imaging approach, with which we demonstrate an application where a photoluminescent sample hidden behind four-layers of opaque polythene films is imaged with a modified multi-photon excitation microscopy setup.
\end{abstract}


\maketitle

\section{Introduction}
The ability to look through random media is important in various fields ranging from life sciences \cite{Gibson2005TOPICAL} to nano-technology \cite{Koenderink2005Optical}. Among all the means, imaging with visible light offers better resolution than microwave and ultrasound imaging and is still considered safe for samples in most scenarios, such as biomedical tissues. In optically-thick random media, wavefront distortions due to multiple light scattering are highly complicated, so the conventional correction methods in adaptive optics \cite{1998aoat.book.....H} provide limited improvement\cite{Tang2012Superpenetration}, which only compensate for low spatial frequency perturbations such as atmospheric turbulence and aberrations induced in translucent tissues \cite{Rueckel2006Adaptive}.
One successful approach in strongly scattering medium is to selectively detect ballistic photons that travel through the medium in a straight line\cite{Hee1993Femtosecond,Niedre2008Early,huang1991optical}. Unfortunately, its working depth is as shallow as a few mean free paths (MFPs) because the ballistic light flux is exponentially extinguished as the depth increases. Photons that are scattered only once\cite{Kang2015Imaging} or slightly so that the overall propagation direction is maintained, such as photo-acoustic microscopy\cite{Xu2011Time,wang2012photoacoustic,Liu2015Optical}, can also be exploited to image target objects located deeper than ten MFPs with limited resolution.

 There are also numbers of other innovative strategies which use scattered waves in full. Pioneering works\cite{Popoff2010Image,Tang2012Superpenetration,Edrei2016Optical,Vellekoop2007Focusing,Vellekoop2010Exploiting,Katz2012Looking,Vellekoop2008Universal,Hsieh2010Imaging,Wang2015Focusing} utilizing the spatial degrees of freedom of light have been demonstrated including wavefront shaping\cite{Vellekoop2007Focusing,Vellekoop2010Exploiting,Katz2012Looking} and phase conjugation\cite{Vellekoop2008Universal,Hsieh2010Imaging,Wang2015Focusing}, which are usually combined with a remarkable speckle correlation phenomenon known as the `memory effect'\cite{Feng1988Correlations,Freund1988Memory} for imaging behind tens of micron- to millimeter- thick tissues. These techniques require the presence of a bright point-source in the object plane and the use of a spatial light modulator or a digital mirror device to generate the needed phase map. Recently, noninvasive optical imaging strategies which remove those requirements have been proposed and demonstrated\cite{Bertolotti2012Non,Katz2014Non}, which exploited the speckle correlations caused by the `memory effect' for scattering and reconstructed the image by numerical methods. These strategies are similar to the speckle interferometry first proposed by Labeyrie\cite{Labeyrie1970Attainment} in 1970,  which aims to yield the diffraction-limited autocorrelation of astronomical objects in spite of atmospheric turbulence. However, in these methods, the image-retrieval process is undependable and time-consuming with stagnation problems. Moreover, the field of view (FOV) is limited by the memory-effect range. Thus, present noninvasive methods are still far from reality. 

 To overcome the above difficulties, we propose that, when a strong scattering system contains sufficient random scatterers to satisfy the ergodic-like condition\cite{Freund1990Looking}, the wave distortions  can be effectively characterized  in the Fourier frequency domain by the phase of the optical-transfer-function, rather than using the full PSF. Once the OTF's phase is obtained, diffraction-limited images can be analytically deconvolved. Based on this theoretical perspective and the analytical solution, we develop a method named Multi-frame OTF Retrieval Engine (MORE), which exploits the redundant information from multiple speckle patterns and rapidly retrieve a faithful OTF phase within several iterative loops. MORE addresses the unreliability and the high computational cost of image restoration in conventional methods. In our experiments, it  starts to converge to the correct solutions from the first iterative loop. The estimated OTF phase as well as images of high-fidelity are achieved after five to eight iterative loops (within 0.2 seconds). Moreover, it can be applied to imaging under weak illumination that is inaccessible with previous phase retrieval methods.  MORE stably yields unique solutions in spite of the low contrast of the camera images and inaccurate initial probes.  
 

With the proposed method, we  demonstrated video imaging at $25\sim 200\,Hz$  of a moving object  hidden behind an opaque layer. Our method does not require any calibration or pre-process. It starts with capturing all the speckle patterns at once, and  then performs post-reconstructions for the images. Furthermore, when combined with a scan of the illuminating source, our technique enables wide-field imaging without limitations on the FOV and the complexity of the object, which used to be two major obstacles in speckle-correlation-based imaging methods.  We also demonstrate an application of this method and obtain a wide-field image of a photoluminescent object hidden behind multiple layers of opaque films, using a modified multi-photon excitation microscopy setup.


\section{physical mechanism}

\begin{figure}[hbt]
    \centering
    \includegraphics[width=80mm]{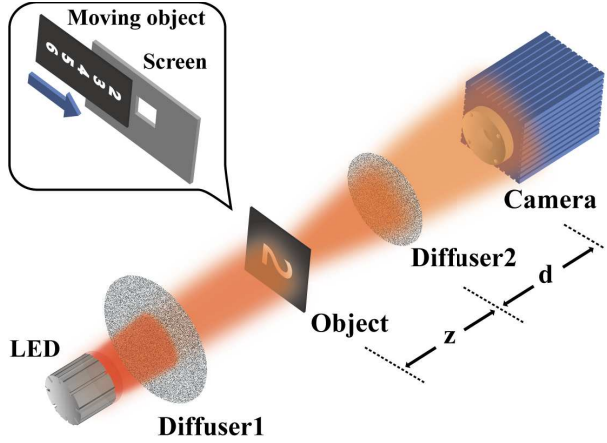}
    \caption{\label{fig:SetupTTM} {Schematic of an imaging system for a time-varying object hidden inside turbid media.} Each diffuser is a piece of 220-grit ground glass. The distance from the object to Diffuser2 is $z\approx 200\;mm$. The distance between Diffuser2 and the camera is $d\approx 300\; mm$. The diameter of Diffuser2 is $\sim6\,mm$. }
    \end{figure}

Figure \ref{fig:SetupTTM} depicts the schematic of our experiment for video imaging a dynamic object hidden between two opaque diffusers which are ground glasses. As shown in the inset, the time-varying object is simulated by a moving object plate in front of a screen with a rectangular aperture. The object plate is a part of a resolution chart board, which has transparent numbers `2' to `6' on it with an average size of $\sim 0.45\,mm\times0.3\,mm$. The plate is driven by a motorized stage, so the content within the rectangular aperture will vary from `2' to `6'. The wavelength of a LED source is $633\, nm$ with a bandwidth of $\sim 15\;nm$, which illuminates the object through Diffuser1. The transmitted light from the object is scattered by Diffuser2, generating a speckle-like intensity pattern which is then detected by a camera. Within a `memory effect' range, spherical waves from nearby points on the object suffer similar distortions and produce shifted but almost identical speckle patterns, $S(x,y)$, on the image plane, which is  the PSF of the system. The detected intensity $I(x,y)$ on the camera is a convolution of the object distribution, $O$, and the PSF:  $I(x,y)= [O{\ast} S](x,y)$. Or, in the frequency domain,
\begin{align}\label{eq:GenImgingFD}
        \tilde{I}(u,v)&= \tilde{S}(u,v) {\cdot}\tilde{O}(u,v)\\ \nonumber
        &= e^{i\Phi_S(u,v)}\cdot |\tilde{I}(u,v)|e^{i\Phi_O(u,v)},
\end{align}
where the tilde denotes the two-dimensional Fourier transform, $u$ and $v$ are spatial frequency coordinates, $\tilde{S}(u,v)$ is the OTF of the system. $\Phi_S(u,v)$ is the OTF's phase. $\Phi_O(u,v)$ is the Fourier phase of the object. The modulus version of eq.(\ref{eq:GenImgingFD}) is $| \tilde{I}(u,v) | =  | \tilde{S}(u,v) | \cdot | \tilde{O}(u,v) | $.
It implies that the magnitude of the OTF acts as 
a spatial frequency filter. $| \tilde{I}(u,v) | $ can be read as the Fourier magnitude of the image formed with the spatial spectrum of the object, $|\tilde{O}(u,v)|$, being filtered by $|\tilde{S}(u,v)|$. 

In a scattering system with enormous numbers of random scatterers, when more than thousands of speckles are captured by the camera's sensor to satisfy the ergodic-like condition\cite{Freund1990Looking}, the magnitude of the OTF can be derived as (see supplemental material\cite{supplemain}),
\begin{equation}\label{eq:PSF_FD}
    |\tilde{S}(u,v)|=\bar{S}\cdot \left[\mathbb{C}^2[T\star T](u,v)+ \delta_D \right]^{1/2},
\end{equation} 
where $T(\xi,\eta)$ represents the distribution of the transmission rate of the scattering medium. $\bar{S}$ is the average intensity of the PSF. $\mathbb{C}$ is the contrast of the PSF. The above equation suggests that the OTF's magnitude of a scattering system is proportional to the square root of the autocorrelation of $T(\xi,\eta)$, except at the origin (0,0). Note that $\delta_D$ is a delta-like peak. 

In contrast, a scattering- and aberration-free lens system has a real OTF, $\tilde{S}_{sf}$, which is proportional to the autocorrelation of the transmission rate of the pupil function. Since  the Fourier transform of the image formed in the lens system, $\tilde{I}_{sf}(u,v)$, also satisfies 
\begin{equation}\label{eq:SF}
    \tilde{I}_{sf}(u,v)= \tilde{S}_{sf}(u,v) {\cdot}\tilde{O}(u,v),
\end{equation} 
its phase is the same as the Fourier phase of the object, i.e., $\Phi_{I_{sf}}(u,v)=\Phi_O(u,v)$. 
  
Therefore, using $| \tilde{I}(u,v) | $ to reconstruct the lost phase information, as in previous noninvasive methods\cite{Bertolotti2012Non,Katz2014Non}, is equivalent to finding an image formed in an aberration-free lens system with its OTF equal to $|\tilde{S}(u,v)|$. After substituting eq.(\ref{eq:GenImgingFD}) and $\tilde{S}_{sf}=|\tilde{S}(u,v)|$ into eq.(\ref{eq:SF}), we obtain  
\begin{equation}\label{eq:SFOTF}
    \tilde{I}_{sf}(u,v)= |\tilde{I}(u,v)|e^{i\Phi_O(u,v)}=\tilde{I}(u,v)e^{-i\Phi_S(u,v)}.
  \end{equation} 
We can see that, the only difference between $\tilde{I}_{sf}(u,v)$ and $\tilde{I}(u,v)$ formed in these two systems is the OTF's phase of the scattering system. Once the OTF's phase is known, a diffraction-limit image can be analytically computed at any time as below, 
\begin{equation}\label{eq:OTFRecoverImage}
    I_{sf}(x,y;t)=\mathcal{F}^{-1}\left\{\tilde{I}(u,v;t)e^{-i\Phi_S(u,v)}\right\},
\end{equation}
where $I(x,y;t)$ is the captured speckle pattern at time $t$. $\mathcal{F}^{-1}$ denotes the inverse Fourier transform. 

This approach preserves the diffraction-limited information determined by the pupil shape $T(\xi,\eta)$. With an even transmission rate over the circular-shape diffuser, the system has a best spatial-frequency resolution $f_{upper} \sim  \frac{nD}{\lambda d}$ at the image plane\cite{Korff1973Analysis} (see also Supplementary Fig. S1\cite{supplemain}), where $n$ denotes the refractive index behind the turbid medium, $D$ is the diameter of the illuminated area on the medium, $\lambda$ is the wavelength. We should mention that, in the object plane, the best spatial-frequency resolution is $\sim  \frac{nD}{\lambda z}$.

\begin{figure}[hbt]
    \centering
    \includegraphics[width=164mm]{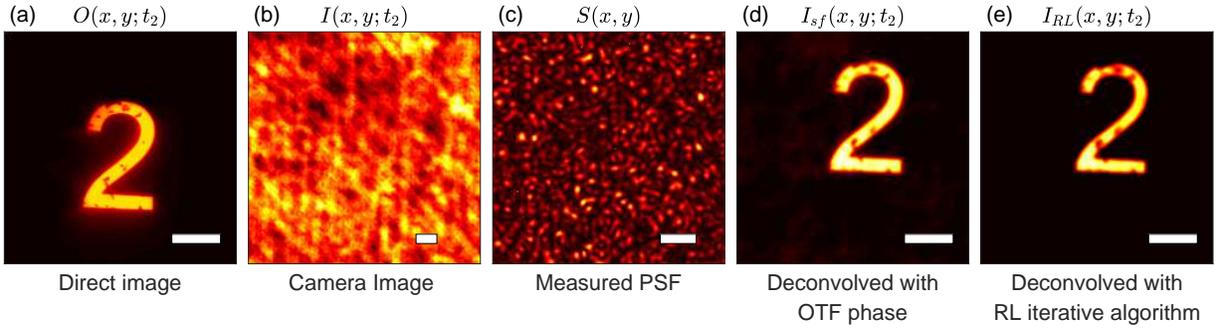}
    \caption{{Image reconstruction from a OTF's phase versus that from a PSF.} (a) The direct image of `2'. (b) The raw camera-captured speckle pattern. (c) The PSF measured with a point source. (d) The image directly deconvolved with the measured OTF phase in (c). (e) The image restored by deconvolution of the speckle pattern and the full PSF with the Richardson-Lucy iterative algorithm (150 iterations). Scale bars: $200\ \mu m$.} \label{fig:OTFvsPSF}
\end{figure} 

As shown in Fig.\ref{fig:OTFvsPSF}, we perform an experimental study by capturing the speckle pattern at $t_2$ while the object is ``2'', and directly measuring the PSF with a $5\;\mu m$ point source. With eq.(\ref{eq:OTFRecoverImage}), the image is deconvolved from the OTF's phase, as shown in Fig.\ref{fig:OTFvsPSF}(d). In Fig.\ref{fig:OTFvsPSF}(e), the  image is retrieved using the Richardson-Lucy algorithm \cite{Richardson1972Bayesian,Lucy1974An} (150 iterations), which is a widely used deconvolution method to construct an image from a known PSF. 
The peak signal-to-noise ratios with the direct image as a reference are $21.7\,\text{dB}$ for Fig.\ref{fig:OTFvsPSF}(d) and $17.9\,\text{dB}$  for Fig.\ref{fig:OTFvsPSF}(e). It appears that, for a strongly scattering system as we have here, the OTF phase itself contains enough information for recovering the image when lacking specific noise or source information. 

\section{Noninvasive imaging for a dynamic object}

For a noninvasive imaging system, the OTF's phase can be obtained in the following way: First, use phase retrieval algorithms to recover the phase information of the object at time $t_0$, e.g. $\Phi_O(u,v;t_0)$; Then, the OTF's phase is directly calculated by $\Phi_S(u,v)=\Phi_I(u,v;t_0)-\Phi_O(u,v;t_0)$. After that, the image of a time-varying object at any time can be computed with eq.(\ref{eq:OTFRecoverImage}). 

\begin{figure}[hbt]
    \centering
    \includegraphics[width=100mm]{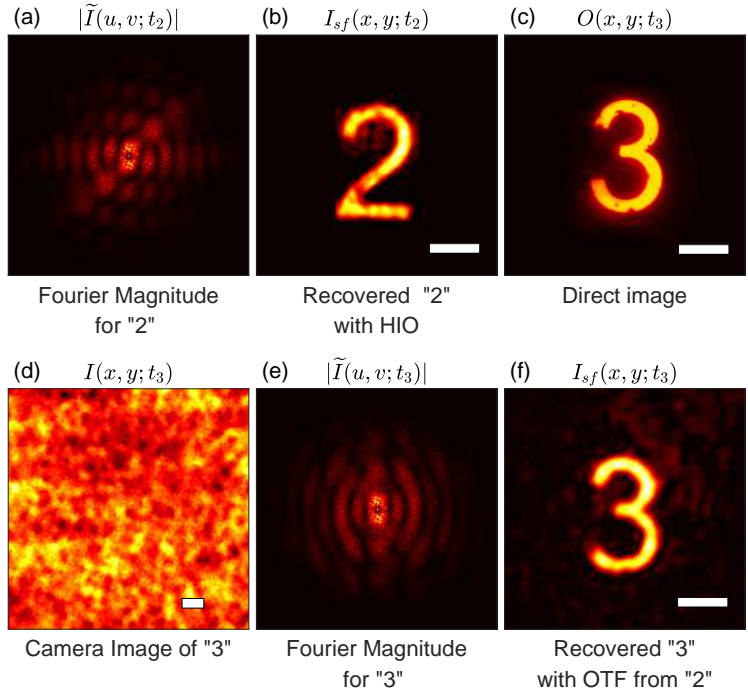}
    \caption{{Recovery of a dynamic object for imaging through a scattering layer using the OTF phase retrieved from a single frame.}  (a) \& (e) Fourier magnitudes of $I(x,y;t_2)$ and $I(x,y;t_3)$, respectively. (b) The reconstructed image from (a) using 'HIO' combined with 'ER'. (c)  The direct images of  `3'. (d)  Raw camera images of `3'. (f) Recovered image of `3' with the OTF's phase retrieved from the speckle pattern for `2'. The average FWHM of speckle grains is $\sim32\,\mu m$, resulting in a resolution in the object plane of $\sim23um$. The theoretical diffraction limit is $\sim 21\,\mu m$. Scale bars: $200\mu m$.} \label{fig:OTFProcess}
\end{figure} 

Based on the above mechanism, we performed the experiment for noninvasive dynamic imaging through turbid media, as shown in Fig.\ref{fig:SetupTTM}. While the object was moving, a sequence of speckled images were recorded by the camera with a $500\,ms$ exposure time for each frame. We selected the first frame and recovered the phase information of the object with the widely used `Hybrid Input-Output (HIO)' combined with `Error-Reduction (ER)' algorithms\cite{Fienup1982Phase},   
as shown in Fig.\ref{fig:OTFProcess}{(b)}. Note that the first frame happens to be of object `2', so we use $t_2$ to denote it. The OTF's phase was then determined as $\Phi_S(u,v)=\Phi_I(u,v;t_2)-\Phi_O(u,v;t_2)$. The images of the rest frames were directly computed with Eq.(\ref{eq:OTFRecoverImage}), which took several milliseconds for each frame. Eventually, a movie for the dynamic object was produced. Fig.\ref{fig:OTFProcess}{(f)} shows the image of $O(x,y;t_3)$ when the object is `3'. 


\section{OTF retrieval from multiple speckle patterns}


However, there are many limits in phase retrieval algorithms from one single speckle pattern. The iterative algorithm is easy to stagnate into a state of local minimum and usually needs many reruns before a satisfying result is selected manually \cite{Bertolotti2012Non,Katz2014Non}. It brings lots of uncertainty to the reconstruction process and makes it time-consuming. The algorithm is even more difficult to converge for complex-shaped object, largely due to the limited information provided from a single pattern, as well as lower camera image contrast when the bright area of the object increases\cite{Katz2014Non}. It appears reasonable that using more speckle patterns resulted from the same or similar scattering area in the calculation will be helpful. These patterns can be simply acquired by illuminating a different part of the object at a time. Or, for dynamic objects such as living biological tissues, where a sequence of camera images are taken to investigate how the sample evolves, the data are readily available. 

\begin{figure}[hbt]
    \centering
    \includegraphics[width=100mm]{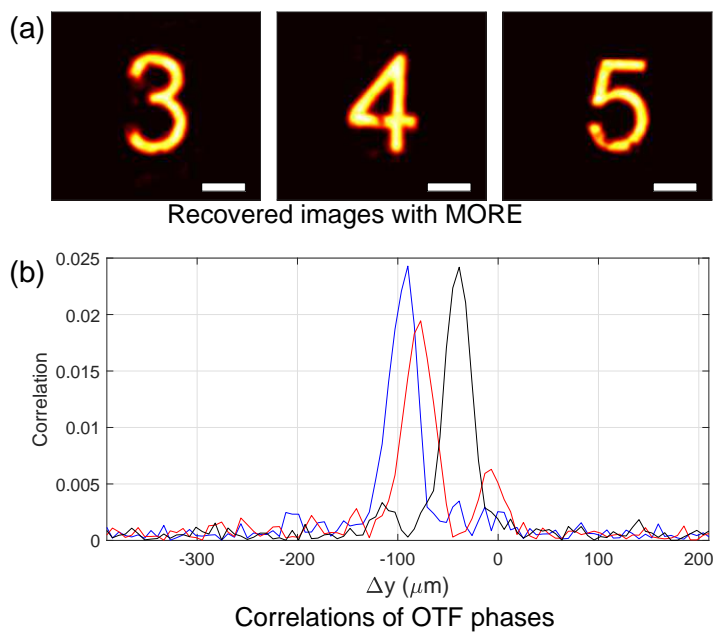}
    \caption{{Performance of MORE.} (color) (a) The recovered images of `3', `4' and '5' obtained in the iterative process of MORE. (b) The correlations between the measured OTF's phase and the calculated ones. The blue curve shows the correlation (along y-axis) between the measured OTF phase and the one derived with MORE; the red and black curves represent the correlations between the measured OTF phase and  those calculated with the speckle patterns for `2' and `3', respectively. Scale bars: $200\mu m$.}  \label{fig:MOREProcess}
\end{figure} 

Here, we propose an iterative approach termed Multi-frame OTF Retrieval Engine (MORE) to take advantage of the redundancy in the measurements. This method is developed in the spirit of the local isoplanatism effects in speckle imaging, i.e., these speckle patterns are generated from different objects but from the same OTF. Usually, two to five speckle patterns (called frames) generated within the same isoplanatic patch are enough for the OTF phase retrieval purpose. MORE consists of several big iterative loops. It starts with an initial guess for the OTF phase, chosen as a random pattern. In one iterative loop, each frame is used for phase retrieval in turn: the realness and non-negativity constraints are used to update the image and the OTF estimate is then renewed by the updated image. From one loop to another, the  OTF phase is kept updated until the iterative process ends. Finally, both the best-estimated images and the OTF phase are recovered simultaneously and robustly. The framework of this algorithm is given in Supplementary Fig. S5\cite{supplemain}. With MORE, the OTF phase is usually retrieved within eight loops. The calculation time is less than 0.2 seconds with a Matlab code running on a Nvidia Tesla P100 GPU.  

For demonstration purpose, we selected three speckle patterns captured when `3', `4' and '5' appeared separately. The recovered images obtained in the iterative process of MORE are shown in Fig.\ref{fig:MOREProcess}(a). With the computed OTF phase from the same process, a movie of the moving object plate can be obtained in the spatial domain using equation (\ref{eq:OTFRecoverImage}).

With MORE, we have reconstructed three videos (400 frames each) of the moving object hidden between two turbid layers at $25\,Hz$ with a $40\,ms$ exposure time (Supplementary Movie 1\cite{supplemovie1}), $100\,Hz$ with a $10\,ms$ exposure time (Supplementary Movie 2\cite{supplemovie2}) and $200\,Hz$  with a $5\,ms$ exposure time (Supplementary Movie 3\cite{supplemovie3}), respectively. The OTF phases are derived for each video individually. Note that, the resolution test board we used in Fig.\ref{fig:OTFvsPSF}, \ref{fig:OTFProcess} and \ref{fig:MOREProcess} was stained. It was substituted by a new and clean one in later experiments, including the videos and the wide-field imaging in the next section.

To investigate the performance of different algorithms, we looked into the correlations between the calculated OTF phases and that obtained by the directly measuring PSF in Fig.\ref{fig:OTFvsPSF}(c). The correlation between OTF phases are computed as the inverse Fourier transform of $e^{i \cdot \{\Phi_{S_a}(u,v)- \Phi_{S_{b}}(u,v)\}}$, which is a real number. $\Phi_{S_{a}}(u,v)$ and $\Phi_{S_{b}}(u,v)$ denote the two OTF phases in comparison. If the two phases are exactly same, the correlation curve will be a $\delta$-like function at the origin. As shown in Fig.\ref{fig:MOREProcess}(b), the blue curve represents the correlation between the measured OTF phase and that obtained by MORE. The red and black curves are for the correlations between the measured OTF phase and those derived from the speckle patterns for `2' and for `3', respectively. The peaks of the blue and black curves are of similar height, and are more than  $20\%$ higher than that of the red curve, suggesting that  MORE and the phase retrieval for the frame of `3' provide better estimates of the OTF phase. However, the reconstruction of `3' with a single frame took more than 20000 iterations in our computation and required  manually selection for the best image. Whereas, MORE converged within 6 loops (equivalent to 18 separate iterations). Note the shifts between the peaks and the origin are attributed to the trivial ambiguities in the phase retrieval problem: it is impossible to determine the absolute position without any prior information. The widths of the  peaks are caused by the diffraction limit of the system.

\begin{figure}[hbt]
    \centering
    \includegraphics[width=90mm]{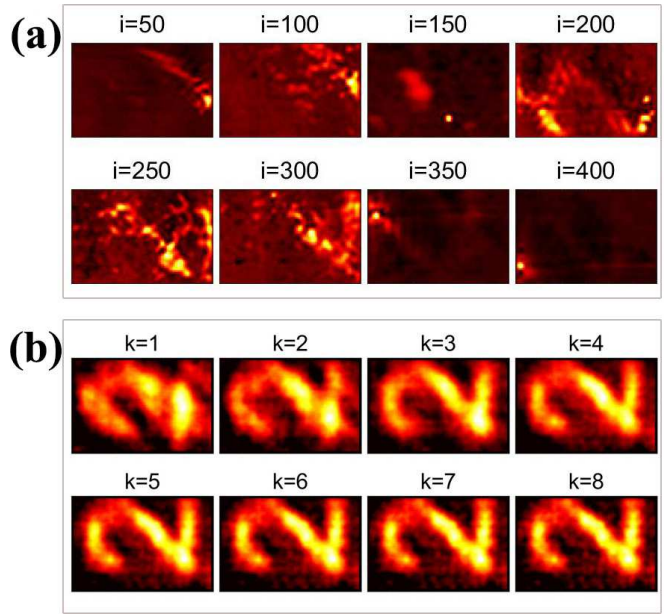}
    \caption{{Performance of the two methods for speckle patterns captured at $200\,Hz$ (with a $5\,ms$ exposure time) for an object hidden behind a diffuser.} (a) Recovered images of iterations ranging from 50 to 400 with the HIO algorithm with one single frame. `i' indicates the iteration number. (b) Recovered images of the big iterative loops with MORE. In this case, three speckle patterns are used in each big loop, which is equivalent to three iterations in the HIO.  `k' is the big loop number. } \label{fig:cycles_200fps}
\end{figure} 

A major advantage of MORE is that it is always stable and provides high-fidelity reconstruction in spite of the low contrast of data and inaccurate initial probes. For the $200\,Hz$ video (with a 5\,$ms$ exposure time), the speckle patterns are corrupted by experimental noise, yielding image contrasts of   only $\sim 7\%$. The method using a single frame failed to give an acceptable reconstruction even after thousands of iterations with various initial conditions. As shown in Fig.\ref{fig:cycles_200fps}(a), the first 400 iterations do not show any sign of convergence, with the real-space image oscillating as a function of the iteration number. In contrast, MORE stably  converges to an unique solution from the first big iterative loop, and provides a reasonable image within five iterative loops, as shown in Fig.\ref{fig:cycles_200fps}(b).   
  
\section{Wide-field imaging through turbid layers}

\begin{figure}[hbt]
    \centering
    \includegraphics[width=130mm]{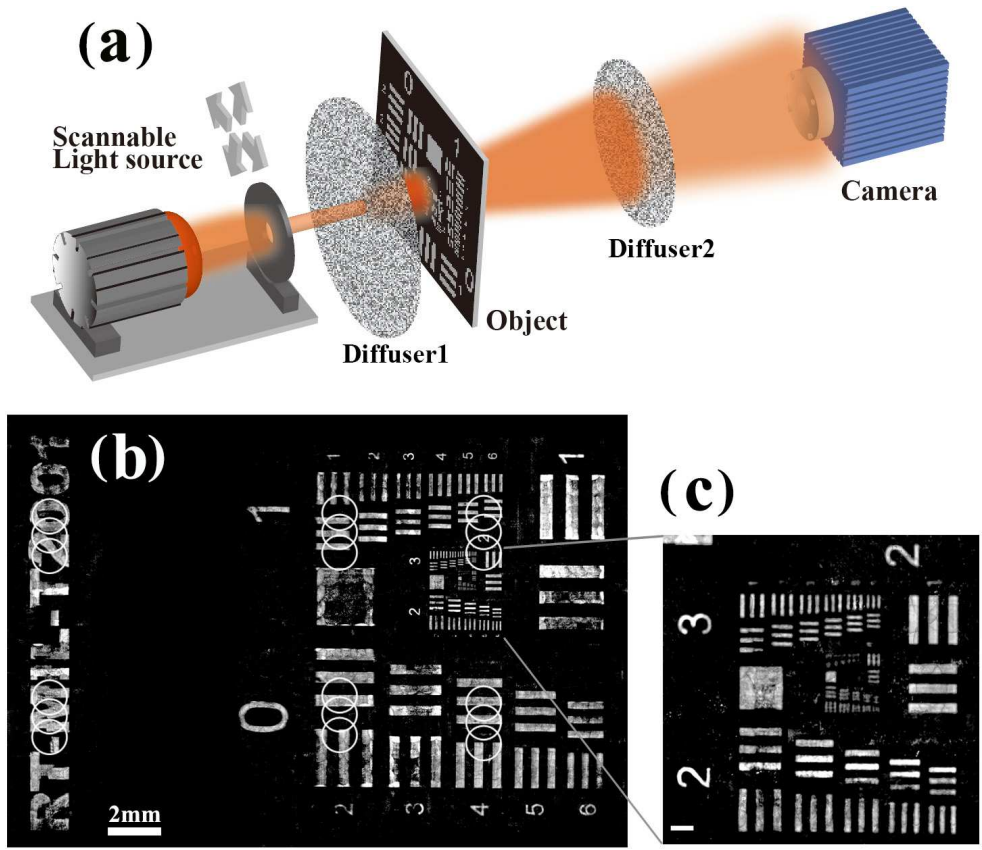}
    \caption{{Wide-field imaging of a complex object through a scattering layer.} (a) Schematic of the experimental setup. (b) Object recovered using MORE. Six sets of three white circles indicate the selected areas used to reconstruct the OTFs by MORE, which are used to recover the whole object. Scale bar: $2\,mm$. (c) An enlargement of the central part of the image. Scale bar at the lower-left corner: $300\,um$. The object plate is a resolution test board (RT-MIL-T2001). Similar to Fig.\ref{fig:SetupTTM}, $z\approx 200\;mm$, $d\approx 300\;mm$ and $D\approx 6\;mm$.} \label{fig:LargObj}
\end{figure}

We emphasize here that, MORE not only provides great advantages in the retrieval processes by largely improving the convergence speed (10- to 1000- fold faster in terms of iterations) and the image quality, but more importantly, it provides solutions with the correct relative positions of all the objects. When combined with a scan of the illuminating source, MORE enables wide-field imaging without the difficulty in aligning multiple pieces of the object before creating the complete picture. This also addresses another drawback in conventional methods, i.e., the difficulty for objects of high complexity\cite{Katz2014Non}.
To demonstrate how to implement wide-field imaging with MORE, we tested it on a large and complex sample, which is a $\sim 27\,mm\times16\,mm$ region on a resolution test board. It also helps to reveal the resolution of this approach. Note that the available FOV with speckle correlations in this configuration (the memory-effect range limited by our detection system) is $\sim 9\,mm\times9\,mm$ in the object plane (Supplementary Fig. S3\cite{supplemain}).  The experimental setup is shown in Fig.\ref{fig:LargObj}(a). The object is hidden between two diffusers. Distances from the object to the front diffuser (Diffuser1) and to the back diffuser (Diffuser2) are $2\,mm$ and $200\,mm$, respectively. We built a scannable light source with a green ($533\, nm$) LED lamp and a $\sim 500\,\mu m$ pinhole, restricting the size of the illuminated area on the object to $\sim \,1.2\,mm$. We scanned the light source over the front diffuser and recorded camera images at each position of the pinhole with an exposure time of $50\,ms$.

Because the wavefront distortions vary at different areas, to acquire a wide-field image of the sample, the phase compensation needs to be determined at each region with a size limited by the memory-effect range. Here, each local OTF phase is retrieved from three speckle patterns taken from three adjacent positions on the object (shown as white circles in Fig.\ref{fig:LargObj}(b) ). A total of six OTFs were computed, which took $\sim 2$ seconds to acquire. Piecing together all the reconstructed parts calculated with the corresponding OTF phases gave us a full view of the object, as shown in Fig.\ref{fig:LargObj}(b) and \ref{fig:LargObj}(c). The smallest feature that can be resolved in our image is Group 4 Element 5 in the resolution test chart, corresponding to an effective resolution of $2^{-(Group+(Element-1)/6+1)}\,mm \approx 20\,\mu m$, in agreement with the expected diffraction-limited resolution of $z\lambda /{ nD}\approx 18\,\mu m$.

\section{Demonstration of an application }
\begin{figure}[hbt]
    \centering
    \includegraphics[width=150mm]{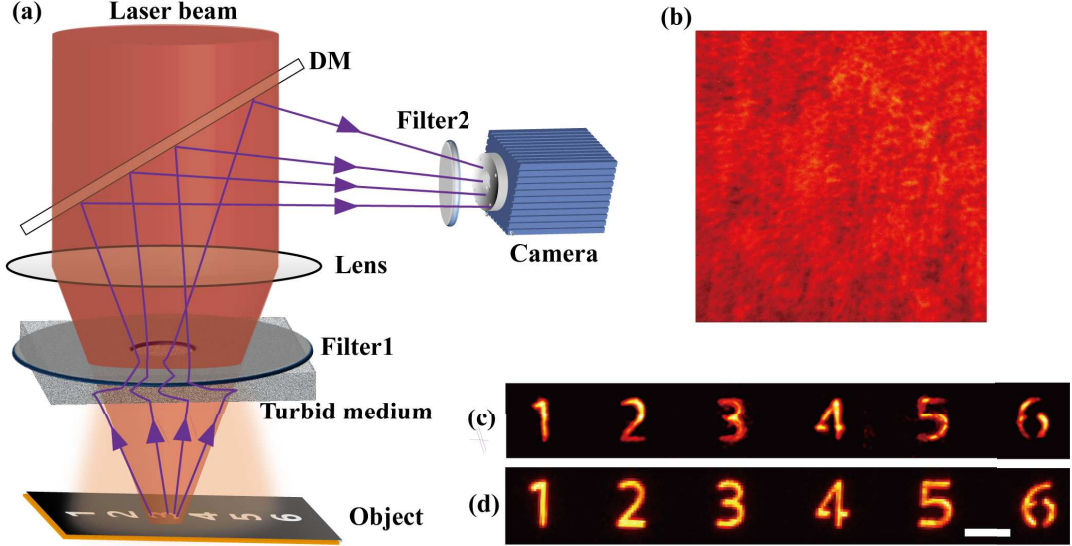}
    \caption{(a) Schematic of wide-field imaging a photoluminescent object through turbid media. DM is a long-pass dichroic mirror that transmits near-infrared light but reflects visible light. Filter2 is a narrow-band pass filter with a bandwidth of $10\;nm$ and a central wavelength at $670\;nm$. The distance between the turbid medium and the object is $10\;cm$. The effective distance from Filter1 to the camera (via DM) is $\sim 10\;cm$. (b) The captured speckle pattern of `3'. (c) The recovered image with MORE. (d) The direct image. The scale bar is $500\;\mu m$. } \label{fig:TPE}
\end{figure}

To achieve wide-field imaging, the requirements for localized illumination within the memory effect range can be easily met if the object is close to or attached to one diffusive layer, as demonstrated above. Another way to circumvent this difficulty for localized illumination/emission is to employ the nonlinear effect of multiphoton-induced fluorescence. For example, in two-photon excitation (TPE), localization of excitation can be sustained both laterally and axially, which can be as small as several microns \cite{Zipfel2003Nonlinear} even with a scattering path of more than 20 MFPs\cite{Gan2000Spatial,Mar1998Monte} for the induced fluorescence. It is attributed to two features: (1) the incident light is at a longer wavelength in the infrared which suffers much weaker scattering (with a MFP more than six times longer) than the induced fluorescence\cite{Gan2000Spatial}; (2) the excitation rate depends non-linearly on the incident intensity, which further localized the fluorescence emission to the focal region. 

We here demonstrate an experiment of imaging a photoluminescent object hidden behind opaque layers, as shown in Fig.\ref{fig:TPE}. The object is prepared by placing a thin plate with hollow numbers `1' to `6' onto a rare-earth upconversion nanophosphors material. The turbid medium consists of four opaque polythene thin films (cut from purple trash bags). The thickness of each film is $\sim 40\;\mu m$. The four films are loosely stacked together with a thin air gap between every two layers. The illumination source is a laser beam at $1550\;nm$ wavelength, which suffers much weaker scattering through the films than the visible light. After the laser light transmits through the turbid medium, there is still a lot of infrared ballistic light that makes a focus of a $\sim 700\;\mu m$ diameter on the object plane. In contrast, visible light (we have tested red and green laser beams) will be severely scattered and fail to form a localized illumination on the object. The focused laser excites multiphoton-induced phosphorescent at $\sim 670\;nm$. Right in front of the turbid medium is a long-pass filter ( Filter1), allowing $1550\;nm$ light passing through but blocking the visible light. We drilled a $\sim 4\;mm$ hole in Filter1, so the excited fluorescence scattered by the turbid medium can escape, which is then reflected into the camera by a dichroic mirror. 

By deflecting the laser beam with galvanometers, we can scan the object with a localized illumination of $\sim 700\;\mu m$ in diameter. Under this setup, the `memory effect' range is around $2\;mm$. The average size of the hollow numbers is $\sim 0.5\;mm$. Therefore, the whole object spans approximately three `memory effect' ranges. With MORE and the technique proposed in last section, we retrieved an OTF for every two adjacent numbers, and then pieced the whole image, as shown in Fig.\ref{fig:TPE}{(c)}. 
%

\section{discussion}

Limitations of the speckle-correlation-based imaging method have been discussed in previous works\cite{Katz2014Non}. One restriction is the practical limit on the complexity, $N,$ of an object that can be recovered in one frame. $N$ is defined by the ratio of the object's bright area to the area of the resolution cell\cite{Katz2014Non}. This limitation is reflected by the wide-field image shown in Fig.\ref{fig:LargObj}(b), where the largest square in the resolution chart of a size $\sim  2.2\;mm$ is not well-recovered. Nevertheless, this problem can be ameliorated by reducing either the size of the localized illumination, or the area of the turbid media where light is collected from.

Another restriction is that, in order to obtain high-fidelity images, the illuminated region of the sample each time must be within an isoplanatic patch (memory effect range), as in adaptive optics, not just laterally but also axially. When the camera sensor is large enough, the lateral and axial speckle correlation ranges are $z\lambda /{\pi nL_{eff}}$\cite{Freund1988Memory} and $(2\lambda/{\pi n})(z/D)^2\;$  \cite{Freund1990Looking}, respectively, where  $L_{eff}$ is the effective thickness of the medium. Any illumination originating from outside of this range contributes to noise in the measurement and deteriorate the image quality.  $L_{eff}= L - l_{t}$ if $L$ is larger than the transport mean free path $l_{t}$.  However, in cases where $L< l_{t}$, $L_{eff}$ will be orders of magnitude smaller than the physical thickness $L$, such as in biological tissues \cite{Katz2014Non} and atmospheric turbulence. In biological soft tissues, $l_{t}$ ranges from hundreds to thousands of microns\cite{Cheong1990A} for visible light, and $n$ is $\sim  1.3 - 1.5$ \cite{Jacques2013Optical}, making it possible to implement this method for an imaging depth more than several hundred microns.

To achieve wide-field imaging, the requirements for localized illumination within the memory effect range can be easily met if the object is close to or attached to one diffusive layer, as demonstrated in Fig.\ref{fig:LargObj}. A pinhole can be used to restrict the illuminated area. Potential applications include non-destructive inspection of flaws and objects on the surface of the inner walls inside thin shells or transparent layers coated with thin films. Another way to circumvent this difficulty for localized illumination/emission is to employ the nonlinear effect of multiphoton-induced fluorescence, which localizes the fluorescence emission to the focal region, like what we demonstrated in Fig.\ref{fig:TPE}. 
 
We have shown that imaging through complex media can be achieved by phase correction in the image frequency domain, where scattering is mainly from distorting layers located above the object. With a known OTF phase calculated from one or multiple speckled images, the millisecond-timescale recovery procedure for each frame paves the way for high-speed dynamic imaging, which can be used in probing complex microscopic and macroscopic processes in different fields. We also develop MORE, a reliable and fast technique for retrieving the OTF through scattering layers and demonstrate experimentally how to use it in conjunction with a scan of the light source to image large and complicated samples.  A major advantage of this iterative scheme is that it is always stable and provides high-fidelity reconstruction in spite of the low contrast of the camera images and inaccurate initial probes, which facilitates  efficient imaging in  previously inaccessible scenarios.   

\section{Acknowledgements}
This work was supported by National Natural Science Foundation of China (grant no. 11503020).
\bibliography{MM2}

\begin{thebibliography}{39}%
\makeatletter
\providecommand \@ifxundefined [1]{%
 \@ifx{#1\undefined}
}%
\providecommand \@ifnum [1]{%
 \ifnum #1\expandafter \@firstoftwo
 \else \expandafter \@secondoftwo
 \fi
}%
\providecommand \@ifx [1]{%
 \ifx #1\expandafter \@firstoftwo
 \else \expandafter \@secondoftwo
 \fi
}%
\providecommand \natexlab [1]{#1}%
\providecommand \enquote  [1]{``#1''}%
\providecommand \bibnamefont  [1]{#1}%
\providecommand \bibfnamefont [1]{#1}%
\providecommand \citenamefont [1]{#1}%
\providecommand \href@noop [0]{\@secondoftwo}%
\providecommand \href [0]{\begingroup \@sanitize@url \@href}%
\providecommand \@href[1]{\@@startlink{#1}\@@href}%
\providecommand \@@href[1]{\endgroup#1\@@endlink}%
\providecommand \@sanitize@url [0]{\catcode `\\12\catcode `\$12\catcode
  `\&12\catcode `\#12\catcode `\^12\catcode `\_12\catcode `\%12\relax}%
\providecommand \@@startlink[1]{}%
\providecommand \@@endlink[0]{}%
\providecommand \url  [0]{\begingroup\@sanitize@url \@url }%
\providecommand \@url [1]{\endgroup\@href {#1}{\urlprefix }}%
\providecommand \urlprefix  [0]{URL }%
\providecommand \Eprint [0]{\href }%
\providecommand \doibase [0]{http://dx.doi.org/}%
\providecommand \selectlanguage [0]{\@gobble}%
\providecommand \bibinfo  [0]{\@secondoftwo}%
\providecommand \bibfield  [0]{\@secondoftwo}%
\providecommand \translation [1]{[#1]}%
\providecommand \BibitemOpen [0]{}%
\providecommand \bibitemStop [0]{}%
\providecommand \bibitemNoStop [0]{.\EOS\space}%
\providecommand \EOS [0]{\spacefactor3000\relax}%
\providecommand \BibitemShut  [1]{\csname bibitem#1\endcsname}%
\let\auto@bib@innerbib\@empty
\bibitem [{\citenamefont {Gibson}\ \emph {et~al.}(2005)\citenamefont {Gibson},
  \citenamefont {Hebden},\ and\ \citenamefont {Arridge}}]{Gibson2005TOPICAL}%
  \BibitemOpen
  \bibfield  {author} {\bibinfo {author} {\bibfnamefont {A.~P.}\ \bibnamefont
  {Gibson}}, \bibinfo {author} {\bibfnamefont {J.~C.}\ \bibnamefont {Hebden}},
  \ and\ \bibinfo {author} {\bibfnamefont {S.~R.}\ \bibnamefont {Arridge}},\
  }\href@noop {} {\bibfield  {journal} {\bibinfo  {journal} {Physics in
  Medicine \& Biology}\ }\textbf {\bibinfo {volume} {50}},\ \bibinfo {pages}
  {R1} (\bibinfo {year} {2005})}\BibitemShut {NoStop}%
\bibitem [{\citenamefont {Koenderink}\ \emph {et~al.}(2005)\citenamefont
  {Koenderink}, \citenamefont {Lagendijk},\ and\ \citenamefont
  {Vos}}]{Koenderink2005Optical}%
  \BibitemOpen
  \bibfield  {author} {\bibinfo {author} {\bibfnamefont {A.~F.}\ \bibnamefont
  {Koenderink}}, \bibinfo {author} {\bibfnamefont {A.}~\bibnamefont
  {Lagendijk}}, \ and\ \bibinfo {author} {\bibfnamefont {W.~L.}\ \bibnamefont
  {Vos}},\ }\href {\doibase 10.1103/PhysRevB.72.153102} {\bibfield  {journal}
  {\bibinfo  {journal} {Phys. Rev. B}\ }\textbf {\bibinfo {volume} {72}},\
  \bibinfo {pages} {153102} (\bibinfo {year} {2005})}\BibitemShut {NoStop}%
\bibitem [{\citenamefont {{Hardy}}(1998)}]{1998aoat.book.....H}%
  \BibitemOpen
  \bibfield  {author} {\bibinfo {author} {\bibfnamefont {J.~W.}\ \bibnamefont
  {{Hardy}}},\ }\href@noop {} {\emph {\bibinfo {title} {Adaptive Optics for
  Astronomical Telescopes, by John W Hardy, pp.~448.~Foreword by John W
  Hardy.~Oxford University Press, Jul 1998.~ISBN-10: 0195090195.~ISBN-13:
  9780195090192}}}\ (\bibinfo {year} {1998})\ p.\ \bibinfo {pages}
  {448}\BibitemShut {NoStop}%
\bibitem [{\citenamefont {Tang}\ \emph {et~al.}(2012)\citenamefont {Tang},
  \citenamefont {Germain},\ and\ \citenamefont
  {Cui}}]{Tang2012Superpenetration}%
  \BibitemOpen
  \bibfield  {author} {\bibinfo {author} {\bibfnamefont {J.}~\bibnamefont
  {Tang}}, \bibinfo {author} {\bibfnamefont {R.~N.}\ \bibnamefont {Germain}}, \
  and\ \bibinfo {author} {\bibfnamefont {M.}~\bibnamefont {Cui}},\ }\href@noop
  {} {\bibfield  {journal} {\bibinfo  {journal} {Proceedings of the National
  Academy of Sciences}\ }\textbf {\bibinfo {volume} {109}},\ \bibinfo {pages}
  {8434} (\bibinfo {year} {2012})}\BibitemShut {NoStop}%
\bibitem [{\citenamefont {Rueckel}\ \emph {et~al.}(2006)\citenamefont
  {Rueckel}, \citenamefont {Mackbucher},\ and\ \citenamefont
  {Denk}}]{Rueckel2006Adaptive}%
  \BibitemOpen
  \bibfield  {author} {\bibinfo {author} {\bibfnamefont {M.}~\bibnamefont
  {Rueckel}}, \bibinfo {author} {\bibfnamefont {J.~A.}\ \bibnamefont
  {Mackbucher}}, \ and\ \bibinfo {author} {\bibfnamefont {W.}~\bibnamefont
  {Denk}},\ }\href@noop {} {\bibfield  {journal} {\bibinfo  {journal}
  {Proceedings of the National Academy of Sciences of the United States of
  America}\ }\textbf {\bibinfo {volume} {103}},\ \bibinfo {pages} {17137}
  (\bibinfo {year} {2006})}\BibitemShut {NoStop}%
\bibitem [{\citenamefont {Hee}\ \emph {et~al.}(1993)\citenamefont {Hee},
  \citenamefont {Swanson}, \citenamefont {Izatt}, \citenamefont {Jacobson},\
  and\ \citenamefont {Fujimoto}}]{Hee1993Femtosecond}%
  \BibitemOpen
  \bibfield  {author} {\bibinfo {author} {\bibfnamefont {M.~R.}\ \bibnamefont
  {Hee}}, \bibinfo {author} {\bibfnamefont {E.~A.}\ \bibnamefont {Swanson}},
  \bibinfo {author} {\bibfnamefont {J.~A.}\ \bibnamefont {Izatt}}, \bibinfo
  {author} {\bibfnamefont {J.~M.}\ \bibnamefont {Jacobson}}, \ and\ \bibinfo
  {author} {\bibfnamefont {J.~G.}\ \bibnamefont {Fujimoto}},\ }\href@noop {}
  {\bibfield  {journal} {\bibinfo  {journal} {Optics Letters}\ }\textbf
  {\bibinfo {volume} {18}},\ \bibinfo {pages} {950} (\bibinfo {year}
  {1993})}\BibitemShut {NoStop}%
\bibitem [{\citenamefont {Niedre}\ and\ \citenamefont
  {Ntziachristos}(2008)}]{Niedre2008Early}%
  \BibitemOpen
  \bibfield  {author} {\bibinfo {author} {\bibfnamefont {M.~J.}\ \bibnamefont
  {Niedre}}\ and\ \bibinfo {author} {\bibfnamefont {V.}~\bibnamefont
  {Ntziachristos}},\ }\href@noop {} {\bibfield  {journal} {\bibinfo  {journal}
  {Proceedings of the National Academy of Sciences}\ }\textbf {\bibinfo
  {volume} {105}},\ \bibinfo {pages} {19126} (\bibinfo {year}
  {2008})}\BibitemShut {NoStop}%
\bibitem [{\citenamefont {Huang}\ \emph {et~al.}(1991)\citenamefont {Huang},
  \citenamefont {Swanson}, \citenamefont {Lin}, \citenamefont {Schuman},
  \citenamefont {Stinson}, \citenamefont {Chang}, \citenamefont {Hee},
  \citenamefont {Flotte}, \citenamefont {Gregory}, \citenamefont {Puliafito}
  \emph {et~al.}}]{huang1991optical}%
  \BibitemOpen
  \bibfield  {author} {\bibinfo {author} {\bibfnamefont {D.}~\bibnamefont
  {Huang}}, \bibinfo {author} {\bibfnamefont {E.~A.}\ \bibnamefont {Swanson}},
  \bibinfo {author} {\bibfnamefont {C.~P.}\ \bibnamefont {Lin}}, \bibinfo
  {author} {\bibfnamefont {J.~S.}\ \bibnamefont {Schuman}}, \bibinfo {author}
  {\bibfnamefont {W.~G.}\ \bibnamefont {Stinson}}, \bibinfo {author}
  {\bibfnamefont {W.}~\bibnamefont {Chang}}, \bibinfo {author} {\bibfnamefont
  {M.~R.}\ \bibnamefont {Hee}}, \bibinfo {author} {\bibfnamefont
  {T.}~\bibnamefont {Flotte}}, \bibinfo {author} {\bibfnamefont
  {K.}~\bibnamefont {Gregory}}, \bibinfo {author} {\bibfnamefont {C.~A.}\
  \bibnamefont {Puliafito}},  \emph {et~al.},\ }\href@noop {} {\bibfield
  {journal} {\bibinfo  {journal} {Science (New York, NY)}\ }\textbf {\bibinfo
  {volume} {254}},\ \bibinfo {pages} {1178} (\bibinfo {year}
  {1991})}\BibitemShut {NoStop}%
\bibitem [{\citenamefont {Kang}\ \emph {et~al.}(2015)\citenamefont {Kang},
  \citenamefont {Jeong}, \citenamefont {Choi}, \citenamefont {Ko},
  \citenamefont {Yang}, \citenamefont {Joo}, \citenamefont {Lee}, \citenamefont
  {Lim}, \citenamefont {Park},\ and\ \citenamefont {Choi}}]{Kang2015Imaging}%
  \BibitemOpen
  \bibfield  {author} {\bibinfo {author} {\bibfnamefont {S.}~\bibnamefont
  {Kang}}, \bibinfo {author} {\bibfnamefont {S.}~\bibnamefont {Jeong}},
  \bibinfo {author} {\bibfnamefont {W.}~\bibnamefont {Choi}}, \bibinfo {author}
  {\bibfnamefont {H.}~\bibnamefont {Ko}}, \bibinfo {author} {\bibfnamefont
  {T.~D.}\ \bibnamefont {Yang}}, \bibinfo {author} {\bibfnamefont {J.~H.}\
  \bibnamefont {Joo}}, \bibinfo {author} {\bibfnamefont {J.~S.}\ \bibnamefont
  {Lee}}, \bibinfo {author} {\bibfnamefont {Y.~S.}\ \bibnamefont {Lim}},
  \bibinfo {author} {\bibfnamefont {Q.~H.}\ \bibnamefont {Park}}, \ and\
  \bibinfo {author} {\bibfnamefont {W.}~\bibnamefont {Choi}},\ }\href@noop {}
  {\bibfield  {journal} {\bibinfo  {journal} {Nature Photonics}\ }\textbf
  {\bibinfo {volume} {9}} (\bibinfo {year} {2015})}\BibitemShut {NoStop}%
\bibitem [{\citenamefont {Xu}\ \emph {et~al.}(2011)\citenamefont {Xu},
  \citenamefont {Liu},\ and\ \citenamefont {Wang}}]{Xu2011Time}%
  \BibitemOpen
  \bibfield  {author} {\bibinfo {author} {\bibfnamefont {X.}~\bibnamefont
  {Xu}}, \bibinfo {author} {\bibfnamefont {H.}~\bibnamefont {Liu}}, \ and\
  \bibinfo {author} {\bibfnamefont {L.~V.}\ \bibnamefont {Wang}},\ }\href@noop
  {} {\bibfield  {journal} {\bibinfo  {journal} {Nature Photonics}\ }\textbf
  {\bibinfo {volume} {5}},\ \bibinfo {pages} {154} (\bibinfo {year}
  {2011})}\BibitemShut {NoStop}%
\bibitem [{\citenamefont {Wang}\ and\ \citenamefont
  {Hu}(2012)}]{wang2012photoacoustic}%
  \BibitemOpen
  \bibfield  {author} {\bibinfo {author} {\bibfnamefont {L.~V.}\ \bibnamefont
  {Wang}}\ and\ \bibinfo {author} {\bibfnamefont {S.}~\bibnamefont {Hu}},\
  }\href@noop {} {\bibfield  {journal} {\bibinfo  {journal} {Science}\ }\textbf
  {\bibinfo {volume} {335}},\ \bibinfo {pages} {1458} (\bibinfo {year}
  {2012})}\BibitemShut {NoStop}%
\bibitem [{\citenamefont {Liu}\ \emph {et~al.}(2015)\citenamefont {Liu},
  \citenamefont {Lai}, \citenamefont {Ma}, \citenamefont {Xu}, \citenamefont
  {Grabar},\ and\ \citenamefont {Wang}}]{Liu2015Optical}%
  \BibitemOpen
  \bibfield  {author} {\bibinfo {author} {\bibfnamefont {Y.}~\bibnamefont
  {Liu}}, \bibinfo {author} {\bibfnamefont {P.}~\bibnamefont {Lai}}, \bibinfo
  {author} {\bibfnamefont {C.}~\bibnamefont {Ma}}, \bibinfo {author}
  {\bibfnamefont {X.}~\bibnamefont {Xu}}, \bibinfo {author} {\bibfnamefont
  {A.~A.}\ \bibnamefont {Grabar}}, \ and\ \bibinfo {author} {\bibfnamefont
  {L.~V.}\ \bibnamefont {Wang}},\ }\href@noop {} {\bibfield  {journal}
  {\bibinfo  {journal} {Nature Communications}\ }\textbf {\bibinfo {volume}
  {6}},\ \bibinfo {pages} {5904} (\bibinfo {year} {2015})}\BibitemShut
  {NoStop}%
\bibitem [{\citenamefont {Popoff}\ \emph {et~al.}(2010)\citenamefont {Popoff},
  \citenamefont {Lerosey}, \citenamefont {Fink}, \citenamefont {Boccara},\ and\
  \citenamefont {Gigan}}]{Popoff2010Image}%
  \BibitemOpen
  \bibfield  {author} {\bibinfo {author} {\bibfnamefont {S.}~\bibnamefont
  {Popoff}}, \bibinfo {author} {\bibfnamefont {G.}~\bibnamefont {Lerosey}},
  \bibinfo {author} {\bibfnamefont {M.}~\bibnamefont {Fink}}, \bibinfo {author}
  {\bibfnamefont {A.~C.}\ \bibnamefont {Boccara}}, \ and\ \bibinfo {author}
  {\bibfnamefont {S.}~\bibnamefont {Gigan}},\ }\href@noop {} {\bibfield
  {journal} {\bibinfo  {journal} {Nature Communications}\ }\textbf {\bibinfo
  {volume} {1}},\ \bibinfo {pages} {626} (\bibinfo {year} {2010})}\BibitemShut
  {NoStop}%
\bibitem [{\citenamefont {Edrei}\ and\ \citenamefont
  {Scarcelli}(2016)}]{Edrei2016Optical}%
  \BibitemOpen
  \bibfield  {author} {\bibinfo {author} {\bibfnamefont {E.}~\bibnamefont
  {Edrei}}\ and\ \bibinfo {author} {\bibfnamefont {G.}~\bibnamefont
  {Scarcelli}},\ }\href@noop {} {\bibfield  {journal} {\bibinfo  {journal}
  {Optica}\ }\textbf {\bibinfo {volume} {3}} (\bibinfo {year}
  {2016})}\BibitemShut {NoStop}%
\bibitem [{\citenamefont {Vellekoop}\ and\ \citenamefont
  {Mosk}(2007)}]{Vellekoop2007Focusing}%
  \BibitemOpen
  \bibfield  {author} {\bibinfo {author} {\bibfnamefont {I.~M.}\ \bibnamefont
  {Vellekoop}}\ and\ \bibinfo {author} {\bibfnamefont {A.~P.}\ \bibnamefont
  {Mosk}},\ }\href@noop {} {\bibfield  {journal} {\bibinfo  {journal} {Optics
  Letters}\ }\textbf {\bibinfo {volume} {32}},\ \bibinfo {pages} {2309}
  (\bibinfo {year} {2007})}\BibitemShut {NoStop}%
\bibitem [{\citenamefont {Vellekoop}\ \emph {et~al.}(2010)\citenamefont
  {Vellekoop}, \citenamefont {Lagendijk},\ and\ \citenamefont
  {Mosk}}]{Vellekoop2010Exploiting}%
  \BibitemOpen
  \bibfield  {author} {\bibinfo {author} {\bibfnamefont {I.~M.}\ \bibnamefont
  {Vellekoop}}, \bibinfo {author} {\bibfnamefont {A.}~\bibnamefont
  {Lagendijk}}, \ and\ \bibinfo {author} {\bibfnamefont {A.~P.}\ \bibnamefont
  {Mosk}},\ }\href@noop {} {\bibfield  {journal} {\bibinfo  {journal} {Nature
  Photonics}\ }\textbf {\bibinfo {volume} {4}},\ \bibinfo {pages} {320}
  (\bibinfo {year} {2010})}\BibitemShut {NoStop}%
\bibitem [{\citenamefont {Katz}\ \emph {et~al.}(2012)\citenamefont {Katz},
  \citenamefont {Small},\ and\ \citenamefont {Silberberg}}]{Katz2012Looking}%
  \BibitemOpen
  \bibfield  {author} {\bibinfo {author} {\bibfnamefont {O.}~\bibnamefont
  {Katz}}, \bibinfo {author} {\bibfnamefont {E.}~\bibnamefont {Small}}, \ and\
  \bibinfo {author} {\bibfnamefont {Y.}~\bibnamefont {Silberberg}},\
  }\href@noop {} {\bibfield  {journal} {\bibinfo  {journal} {Nature Photonics}\
  }\textbf {\bibinfo {volume} {6}},\ \bibinfo {pages} {549} (\bibinfo {year}
  {2012})}\BibitemShut {NoStop}%
\bibitem [{\citenamefont {Vellekoop}\ and\ \citenamefont
  {Mosk}(2008)}]{Vellekoop2008Universal}%
  \BibitemOpen
  \bibfield  {author} {\bibinfo {author} {\bibfnamefont {I.~M.}\ \bibnamefont
  {Vellekoop}}\ and\ \bibinfo {author} {\bibfnamefont {A.~P.}\ \bibnamefont
  {Mosk}},\ }\href {\doibase 10.1103/PhysRevLett.101.120601} {\bibfield
  {journal} {\bibinfo  {journal} {Phys. Rev. Lett.}\ }\textbf {\bibinfo
  {volume} {101}},\ \bibinfo {pages} {120601} (\bibinfo {year}
  {2008})}\BibitemShut {NoStop}%
\bibitem [{\citenamefont {Hsieh}\ \emph {et~al.}(2010)\citenamefont {Hsieh},
  \citenamefont {Pu}, \citenamefont {Grange}, \citenamefont {Laporte},\ and\
  \citenamefont {Psaltis}}]{Hsieh2010Imaging}%
  \BibitemOpen
  \bibfield  {author} {\bibinfo {author} {\bibfnamefont {C.~L.}\ \bibnamefont
  {Hsieh}}, \bibinfo {author} {\bibfnamefont {Y.}~\bibnamefont {Pu}}, \bibinfo
  {author} {\bibfnamefont {R.}~\bibnamefont {Grange}}, \bibinfo {author}
  {\bibfnamefont {G.}~\bibnamefont {Laporte}}, \ and\ \bibinfo {author}
  {\bibfnamefont {D.}~\bibnamefont {Psaltis}},\ }\href@noop {} {\bibfield
  {journal} {\bibinfo  {journal} {Optics Express}\ }\textbf {\bibinfo {volume}
  {18}},\ \bibinfo {pages} {20723} (\bibinfo {year} {2010})}\BibitemShut
  {NoStop}%
\bibitem [{\citenamefont {Wang}\ \emph {et~al.}(2015)\citenamefont {Wang},
  \citenamefont {Zhou}, \citenamefont {Brake}, \citenamefont {Ruan},
  \citenamefont {Jang},\ and\ \citenamefont {Yang}}]{Wang2015Focusing}%
  \BibitemOpen
  \bibfield  {author} {\bibinfo {author} {\bibfnamefont {D.}~\bibnamefont
  {Wang}}, \bibinfo {author} {\bibfnamefont {E.~H.}\ \bibnamefont {Zhou}},
  \bibinfo {author} {\bibfnamefont {J.}~\bibnamefont {Brake}}, \bibinfo
  {author} {\bibfnamefont {H.}~\bibnamefont {Ruan}}, \bibinfo {author}
  {\bibfnamefont {M.}~\bibnamefont {Jang}}, \ and\ \bibinfo {author}
  {\bibfnamefont {C.}~\bibnamefont {Yang}},\ }\href@noop {} {\bibfield
  {journal} {\bibinfo  {journal} {Optica}\ }\textbf {\bibinfo {volume} {2}},\
  \bibinfo {pages} {728} (\bibinfo {year} {2015})}\BibitemShut {NoStop}%
\bibitem [{\citenamefont {Feng}\ \emph {et~al.}(1988)\citenamefont {Feng},
  \citenamefont {Kane}, \citenamefont {Lee},\ and\ \citenamefont
  {Stone}}]{Feng1988Correlations}%
  \BibitemOpen
  \bibfield  {author} {\bibinfo {author} {\bibfnamefont {S.}~\bibnamefont
  {Feng}}, \bibinfo {author} {\bibfnamefont {C.}~\bibnamefont {Kane}}, \bibinfo
  {author} {\bibfnamefont {P.~A.}\ \bibnamefont {Lee}}, \ and\ \bibinfo
  {author} {\bibfnamefont {A.~D.}\ \bibnamefont {Stone}},\ }\href@noop {}
  {\bibfield  {journal} {\bibinfo  {journal} {Physical Review Letters}\
  }\textbf {\bibinfo {volume} {61}},\ \bibinfo {pages} {834} (\bibinfo {year}
  {1988})}\BibitemShut {NoStop}%
\bibitem [{\citenamefont {Freund}\ \emph {et~al.}(1988)\citenamefont {Freund},
  \citenamefont {Rosenbluh},\ and\ \citenamefont {Feng}}]{Freund1988Memory}%
  \BibitemOpen
  \bibfield  {author} {\bibinfo {author} {\bibfnamefont {I.~I.}\ \bibnamefont
  {Freund}}, \bibinfo {author} {\bibfnamefont {M.}~\bibnamefont {Rosenbluh}}, \
  and\ \bibinfo {author} {\bibfnamefont {S.}~\bibnamefont {Feng}},\ }\href@noop
  {} {\bibfield  {journal} {\bibinfo  {journal} {Physical Review Letters}\
  }\textbf {\bibinfo {volume} {61}},\ \bibinfo {pages} {2328} (\bibinfo {year}
  {1988})}\BibitemShut {NoStop}%
\bibitem [{\citenamefont {Bertolotti}\ \emph {et~al.}(2012)\citenamefont
  {Bertolotti}, \citenamefont {van Putten}, \citenamefont {Blum}, \citenamefont
  {Lagendijk}, \citenamefont {Vos},\ and\ \citenamefont
  {Mosk}}]{Bertolotti2012Non}%
  \BibitemOpen
  \bibfield  {author} {\bibinfo {author} {\bibfnamefont {J.}~\bibnamefont
  {Bertolotti}}, \bibinfo {author} {\bibfnamefont {E.~G.}\ \bibnamefont {van
  Putten}}, \bibinfo {author} {\bibfnamefont {C.}~\bibnamefont {Blum}},
  \bibinfo {author} {\bibfnamefont {A.}~\bibnamefont {Lagendijk}}, \bibinfo
  {author} {\bibfnamefont {W.~L.}\ \bibnamefont {Vos}}, \ and\ \bibinfo
  {author} {\bibfnamefont {A.~P.}\ \bibnamefont {Mosk}},\ }\href@noop {}
  {\bibfield  {journal} {\bibinfo  {journal} {Nature}\ }\textbf {\bibinfo
  {volume} {491}},\ \bibinfo {pages} {232} (\bibinfo {year}
  {2012})}\BibitemShut {NoStop}%
\bibitem [{\citenamefont {Katz}\ \emph {et~al.}(2014)\citenamefont {Katz},
  \citenamefont {Heidmann}, \citenamefont {Fink},\ and\ \citenamefont
  {Gigan}}]{Katz2014Non}%
  \BibitemOpen
  \bibfield  {author} {\bibinfo {author} {\bibfnamefont {O.}~\bibnamefont
  {Katz}}, \bibinfo {author} {\bibfnamefont {P.}~\bibnamefont {Heidmann}},
  \bibinfo {author} {\bibfnamefont {M.}~\bibnamefont {Fink}}, \ and\ \bibinfo
  {author} {\bibfnamefont {S.}~\bibnamefont {Gigan}},\ }\href@noop {}
  {\bibfield  {journal} {\bibinfo  {journal} {Nature Photonics}\ }\textbf
  {\bibinfo {volume} {8}},\ \bibinfo {pages} {784} (\bibinfo {year}
  {2014})}\BibitemShut {NoStop}%
\bibitem [{\citenamefont {Labeyrie}(1970)}]{Labeyrie1970Attainment}%
  \BibitemOpen
  \bibfield  {author} {\bibinfo {author} {\bibfnamefont {A.}~\bibnamefont
  {Labeyrie}},\ }\href@noop {} {\bibfield  {journal} {\bibinfo  {journal}
  {Astronomy \& Astrophysics}\ }\textbf {\bibinfo {volume} {6}},\ \bibinfo
  {pages} {85} (\bibinfo {year} {1970})}\BibitemShut {NoStop}%
\bibitem [{\citenamefont {Freund}(1990)}]{Freund1990Looking}%
  \BibitemOpen
  \bibfield  {author} {\bibinfo {author} {\bibfnamefont {I.}~\bibnamefont
  {Freund}},\ }\href@noop {} {\bibfield  {journal} {\bibinfo  {journal}
  {Physica A: Statistical Mechanics \& Its Applications}\ }\textbf {\bibinfo
  {volume} {168}},\ \bibinfo {pages} {49} (\bibinfo {year} {1990})}\BibitemShut
  {NoStop}%
\bibitem [{sup({\natexlab{a}})}]{supplemain}%
  \BibitemOpen
  \href@noop {} {} ({\natexlab{a}}),\ \bibinfo {note} {see Supplemental
  Material at SupplementalMaterial.docx for details}\BibitemShut {NoStop}%
\bibitem [{\citenamefont {Korff}(1973)}]{Korff1973Analysis}%
  \BibitemOpen
  \bibfield  {author} {\bibinfo {author} {\bibfnamefont {D.}~\bibnamefont
  {Korff}},\ }\href@noop {} {\bibfield  {journal} {\bibinfo  {journal} {Journal
  of the Optical Society of America}\ }\textbf {\bibinfo {volume} {63}},\
  \bibinfo {pages} {971} (\bibinfo {year} {1973})}\BibitemShut {NoStop}%
\bibitem [{\citenamefont {Richardson}(1972)}]{Richardson1972Bayesian}%
  \BibitemOpen
  \bibfield  {author} {\bibinfo {author} {\bibfnamefont {W.~H.}\ \bibnamefont
  {Richardson}},\ }\href@noop {} {\bibfield  {journal} {\bibinfo  {journal}
  {Journal of the Optical Society of America}\ }\textbf {\bibinfo {volume}
  {62}},\ \bibinfo {pages} {55} (\bibinfo {year} {1972})}\BibitemShut {NoStop}%
\bibitem [{\citenamefont {Lucy}(1974)}]{Lucy1974An}%
  \BibitemOpen
  \bibfield  {author} {\bibinfo {author} {\bibfnamefont {L.~B.}\ \bibnamefont
  {Lucy}},\ }\href@noop {} {\bibfield  {journal} {\bibinfo  {journal}
  {Astronomical Journal}\ }\textbf {\bibinfo {volume} {79}},\ \bibinfo {pages}
  {745} (\bibinfo {year} {1974})}\BibitemShut {NoStop}%
\bibitem [{\citenamefont {Fienup}(1982)}]{Fienup1982Phase}%
  \BibitemOpen
  \bibfield  {author} {\bibinfo {author} {\bibfnamefont {J.~R.}\ \bibnamefont
  {Fienup}},\ }\href@noop {} {\bibfield  {journal} {\bibinfo  {journal}
  {Applied Optics}\ }\textbf {\bibinfo {volume} {21}},\ \bibinfo {pages} {2758}
  (\bibinfo {year} {1982})}\BibitemShut {NoStop}%
\bibitem [{sup({\natexlab{b}})}]{supplemovie1}%
  \BibitemOpen
  \href@noop {} {} ({\natexlab{b}}),\ \bibinfo {note} {see Supplemental
  Material at SupplementaryMovie1.avi for the 25Hz video imaging of a dynamic
  object through a turbid medium}\BibitemShut {NoStop}%
\bibitem [{sup({\natexlab{c}})}]{supplemovie2}%
  \BibitemOpen
  \href@noop {} {} ({\natexlab{c}}),\ \bibinfo {note} {see Supplemental
  Material at SupplementaryMovie2.avi for the 100Hz video imaging of a dynamic
  object through a turbid medium}\BibitemShut {NoStop}%
\bibitem [{sup({\natexlab{d}})}]{supplemovie3}%
  \BibitemOpen
  \href@noop {} {} ({\natexlab{d}}),\ \bibinfo {note} {see Supplemental
  Material at SupplementaryMovie3.avi for the 200Hz video imaging of a dynamic
  object through a turbid medium}\BibitemShut {NoStop}%
\bibitem [{\citenamefont {Zipfel}\ \emph {et~al.}(2003)\citenamefont {Zipfel},
  \citenamefont {Williams},\ and\ \citenamefont {Webb}}]{Zipfel2003Nonlinear}%
  \BibitemOpen
  \bibfield  {author} {\bibinfo {author} {\bibfnamefont {W.~R.}\ \bibnamefont
  {Zipfel}}, \bibinfo {author} {\bibfnamefont {R.~M.}\ \bibnamefont
  {Williams}}, \ and\ \bibinfo {author} {\bibfnamefont {W.~W.}\ \bibnamefont
  {Webb}},\ }\href@noop {} {\bibfield  {journal} {\bibinfo  {journal} {Nature
  Biotechnology}\ }\textbf {\bibinfo {volume} {21}},\ \bibinfo {pages} {1369}
  (\bibinfo {year} {2003})}\BibitemShut {NoStop}%
\bibitem [{\citenamefont {Gan}\ and\ \citenamefont
  {Gu}(2000)}]{Gan2000Spatial}%
  \BibitemOpen
  \bibfield  {author} {\bibinfo {author} {\bibfnamefont {X.}~\bibnamefont
  {Gan}}\ and\ \bibinfo {author} {\bibfnamefont {M.}~\bibnamefont {Gu}},\
  }\href@noop {} {\bibfield  {journal} {\bibinfo  {journal} {Applied Optics}\
  }\textbf {\bibinfo {volume} {39}},\ \bibinfo {pages} {1575} (\bibinfo {year}
  {2000})}\BibitemShut {NoStop}%
\bibitem [{\citenamefont {Mar~Blanca}\ and\ \citenamefont
  {Saloma}(1998)}]{Mar1998Monte}%
  \BibitemOpen
  \bibfield  {author} {\bibinfo {author} {\bibfnamefont {C.}~\bibnamefont
  {Mar~Blanca}}\ and\ \bibinfo {author} {\bibfnamefont {C.}~\bibnamefont
  {Saloma}},\ }\href@noop {} {\bibfield  {journal} {\bibinfo  {journal}
  {Applied Optics}\ }\textbf {\bibinfo {volume} {37}},\ \bibinfo {pages} {8092}
  (\bibinfo {year} {1998})}\BibitemShut {NoStop}%
\bibitem [{\citenamefont {Cheong}\ \emph {et~al.}(1990)\citenamefont {Cheong},
  \citenamefont {Prahl},\ and\ \citenamefont {Welch}}]{Cheong1990A}%
  \BibitemOpen
  \bibfield  {author} {\bibinfo {author} {\bibfnamefont {W.~F.}\ \bibnamefont
  {Cheong}}, \bibinfo {author} {\bibfnamefont {S.~A.}\ \bibnamefont {Prahl}}, \
  and\ \bibinfo {author} {\bibfnamefont {A.~J.}\ \bibnamefont {Welch}},\
  }\href@noop {} {\bibfield  {journal} {\bibinfo  {journal} {IEEE Journal of
  Quantum Electronics}\ }\textbf {\bibinfo {volume} {26}},\ \bibinfo {pages}
  {2166} (\bibinfo {year} {1990})}\BibitemShut {NoStop}%
\bibitem [{\citenamefont {Jacques}(2013)}]{Jacques2013Optical}%
  \BibitemOpen
  \bibfield  {author} {\bibinfo {author} {\bibfnamefont {S.~L.}\ \bibnamefont
  {Jacques}},\ }\href@noop {} {\bibfield  {journal} {\bibinfo  {journal}
  {Physics in Medicine \& Biology}\ }\textbf {\bibinfo {volume} {58}},\
  \bibinfo {pages} {R37} (\bibinfo {year} {2013})}\BibitemShut {NoStop}%
\end{thebibliography}%
\end{document}